\documentclass[reprint,prl]{revtex4-2}
\usepackage{amssymb}
\usepackage{amsmath}
\usepackage{braket}
\usepackage{color}
\usepackage{MnSymbol}
\usepackage[amsmath]{empheq}
\usepackage{tikz}
\usepackage{color}
\usepackage{enumerate}
\usepackage{comment}
\usepackage{nicefrac}
\usepackage{bbm}
\usepackage{tensor}
\usepackage{tikzsymbols}
\usepackage{ifsym}
\usepackage{graphicx}

\newcommand{\be}{\begin{equation}}
\newcommand{\ee}{\end{equation}}
\newcommand{\ba}{\begin{aligned}}
\newcommand{\ea}{\end{aligned}}
\newcommand{\bs}[1]{\boldsymbol{\bf #1}}

\usepackage{cancel}

\begin{document}

\title{
Global Quenches after Localised Perturbations
}
\author{Maurizio Fagotti}
\affiliation{Universit\'e Paris-Saclay, CNRS, LPTMS, 91405, Orsay, France}

\date{\today}

\begin{abstract}
We investigate the effect of a single spin flip preceding a global quench between  translationally invariant  local Hamiltonians in spin-$\frac{1}{2}$ chains.
The effect of the localised perturbation does not fade away however large the distance from the perturbation is. In particular, translational invariance is not restored and the infinite time limit depends on whether the spin was flipped or not. We argue that this phenomenon is more general than the particular example considered and we conjecture that it is triggered by topological properties, specifically, the existence of ``semilocal charges''. 
\end{abstract}
\maketitle

\paragraph{Introduction.}
In its simplest acceptation,  ``quantum quench'' is a modern term indicating the time evolution of the ground state of some physically meaningful Hamiltonian of a quantum many-body system after a sudden change of a Hamiltonian parameter. The quench was dubbed ``global'' (presumably around the appearance of Refs~\cite{Calabrese2005Evolution,Eisler2007Evolution}) when the parameter is coupled to an extensive operator in such a way that the initial state has a significant overlap with an exponentially large number (with respect to the system size) of excited states. Probably the first analytical study of such dynamics in a spin chain dates back to $1970$, with Ref.~\cite{Barouch1970Statistical}. Global quenches have attracted a lot of attention in the last two decades for  being a bridge between nonequilibrium and equilibrium physics~\cite{Polkovnikov2011Colloquium,Gogolin2016Equilibration,Eisert2015Quantum,Essler2016Quench,Vidmar2016Generalized,Calabrese2016Quantum}. A global quench results in complex dynamics, but at long enough time most of the details become negligible and the local properties of the state fit in the domain of statistical theories originally conceived to describe systems at equilibrium\cite{Srednicki1994Chaos,Rigol2012Alternatives}.  
Arguably, most of our intuition about global quenches comes from studies of  translationally invariant systems. Translational invariance significantly reduces complexity, and analytical studies in particularly simple models become feasible. 
In this respect, integrable models have been playing a key role~\cite{korepin_bogoliubov_izergin_1993,Caux2013Time,Bonnes2014Light-Cone,Piroli2017What,Borsi2020Current,Klobas2021Exact,Klobas2021Exact1}, both for providing a clue of which phenomena could be observed in a nonequilibrium quantum many-body system and for the exceptional, exotic physics they personify. 
Integrable systems do not thermalise: they approach generalised Gibbs ensembles (GGE)~\cite{Rigol2007Relaxation,Cramer2008Exact,Calabrese2011Quantum,Fagotti2013Reduced,Ilievski2015Complete,Essler2015Generalized}. Subtleties aside, this is now established, and it has strongly influenced the recent theoretical attempts to relax the hypothesis of translational invariance. Localised defects (see, e.g.,~\cite{Bernard2015Non-equilibrium,Fagotti2016Control,Bertini2016Determination,Ljubotina2019Non,Bastianello2018Nonequilibrium,Bastianello2019Lack}) and more global inhomogeneities (see, e.g.,~\cite{Biella2016Energy,Doyon2017A,Dubail2017Conformal,Cao2018Incomplete,Biella2019Ballistic,Bastianello2019Generalized,Ruggiero2020Quantum, Malvania2021Generalized}) in the Hamiltonian on one side and inhomogeneities in the initial state (see, e.g., ~\cite{Bernard2016Conformal,Bertini2016Transport,Castro-Alvaredo2016Emergent,Piroli2017Transport,Bulchandani2017Solvable,Bulchandani2018Bethe-Boltzmann}) on the other are often explained by establishing a connection with the accepted phenomenon of local relaxation to a GGE. When this approach does not work, it means that something is still not fully understood. For instance, we remind the reader of the failure~\cite{Brockmann2014Quench,Wouters2014Quenching,Pozsgay2014Correlations,Mestyan2015Quenching} of the original formulation~\cite{Fagotti2013Stationary,Pozsgay2013The,Fagotti2014Relaxation} of GGE in the Heisenberg XXZ chain, which unveiled the importance of so-called quasilocal conservation laws~\cite{Ilievski2015Complete,Ilievski2015Quasilocal,Ilievski2016Quasilocal,Piroli2016Exact}; or the failure~\cite{Fagotti2014On} of the GGE constructed with mode occupation numbers in noninteracting chains~\cite{Rigol2007Relaxation}, which unveiled the importance of conservation laws that do not belong to a complete set of charges~\cite{Fagotti2014On,Fagotti2016Charges,Doyon2017Thermalization,Zadnik2016Quasilocal}. 

In this Letter we report another piece of the puzzle that has been overlooked. We consider a  global quench from a very simple initial state that is translationally invariant everywhere except for a finite region. Time evolution is generated by a translationally invariant local spin-chain Hamiltonian. Normally, we would expect the weak translational-symmetry breaking in the initial state to be irrelevant at large times and distances. We show that this is not the case. 

\paragraph{The example.}
We consider a spin chain described by the Hamiltonian
\be\label{eq:Hsigma}
\bs H=\sum\nolimits_{j=1}^L\bs\sigma_{j-1}^x\Bigl(J_x\bs 1-J_y\bs \sigma_j^z\Bigr)\bs \sigma_{j+1}^x+J_z \bs\sigma_j^z\, ,
\ee
where periodic boundary conditions are understood.
This model is dual to the Heisenberg XYZ one, so it is integrable. 
The duality transformation is manifest in terms of the auxiliary Pauli matrices
\be\label{eq:duality}
\bs \tau_j^z= -\prod_{n=j}^{L-1}\bs\sigma_n^z\bs\sigma_L^y\ ,\quad \bs \tau_j^x= \begin{cases}
-\bs\sigma_1^y \prod_{n=2}^{L-1}\bs\sigma_n^z \bs \sigma_L^y,&j=1\\
\bs\sigma_{j-1}^x\bs\sigma_j^x,&j>1\, .
\end{cases}
\ee
The Hamiltonian can indeed be written as follows
\be\label{eq:H}
\bs H=\sideset{}{^\prime}\sum_{j=1}
^L(J_x\bs\tau_j^x\bs\tau_{j+1}^x+J_y\bs\tau_j^y\bs\tau_{j+1}^y+J_z \bs\tau_j^z\bs\tau_{j+1}^z
)
\ee
where we introduced the notation
\be
\sideset{}{^\prime}\sum_{j=1}
^L\bs O_j := \tfrac{\bs 1+\bs \Pi_{\tau}^x}{2}\sideset{}{^+}\sum^L_{j=1
} \!\!\bs O_j \tfrac{\bs 1+\bs \Pi_{\tau}^x}{2}+
 \tfrac{\bs 1-\bs \Pi_{\tau}^x}{2}\bs\tau_1^z \sideset{}{^-}\sum_{j=1
 }^L\!\! \bs O_j \bs\tau_1^z \tfrac{\bs 1-\bs \Pi_{ \tau}^x}{2},
\ee
with
$
\bs \Pi_{\tau}^\alpha= \prod_{j=1}^L\bs \tau_j^\alpha
$ and $\sum\nolimits^\pm$ standing for the sum in which $\bs\tau^{y,z}_{L+1}=\pm \bs\tau^{y,z}_1$. 
Note that the projectors into the distinct sectors can be easily written in terms of the $\bs\sigma$ operators, indeed we have $\bs \Pi_{\tau}^x=\bs \Pi_{\sigma}^z$ and $\bs\tau_1^z \frac{\bs 1-\bs \Pi_{\tau}^x}{2}=i \bs \sigma_L^x\frac{\bs 1-\bs\Pi^z_{ \sigma}}{2}$.
We warn the reader that duality transformations affect the notion of locality and, in turn, the physical properties of a system: \eqref{eq:Hsigma} does not describe the Heisenberg model.

At the initial time we prepare the system in the state 
\be\label{eq:0}
\ket{\Psi(0)}=\ket{\uparrow_1\cdots\uparrow_{L-1}\downarrow_L}_{\sigma}\equiv \ket{\Uparrow\downarrow_L}_\sigma\, ,
\ee
where the subscript $\sigma$ means that we are representing the spins with the Pauli matrices $\bs \sigma$,  $\Uparrow$ ($\Downarrow$) stands for a string of $\uparrow$'s ($\downarrow$'s) of whatever length.
The last spin is flipped to eventually simplify the boundary conditions in the dual representation, but we stress that this is completely irrelevant in the thermodynamic limit, as the Lieb-Robinson bounds~\cite{Lieb1972The} rule out any effect in the bulk, provided that $t\ll L$.
We then flip a spin in the bulk, let's say, at position $1\ll \ell\ll L$. This is achieved by applying $\bs \sigma_\ell^x$ to the state; we then have
\be\label{eq:Psi0}
\ket{\Psi(0^+)}=\ket{\Uparrow\downarrow_\ell\Uparrow\downarrow_L}_{\sigma}\, .
\ee
This state is in the sector with $\bs \Pi_{\sigma}^z=1$, and hence $\bs\Pi^x_{\tau}=1$, and in the $\tau$-representation has the form of a cat state:
\be
\ket{\Psi(0^+)}=\tfrac{1}{\sqrt{2}}(\ket{\Uparrow_\ell\Downarrow}_{\tau}+\ket{\Downarrow_\ell\Uparrow}_{\tau})\, .
\ee
where the subscript $\ell$ indicates that the string ends at $\ell$.
Since $\bs\Pi^x_{\tau}=1$, Hamiltonian \eqref{eq:H} is equivalent to
\be\label{eq:Htau}
\tilde{\bs H}=\sum\nolimits_{j=1}^L \bs (J_x\bs\tau_j^x\bs\tau_{j+1}^x+J_y\bs\tau_j^y\bs\tau_{j+1}^y+J_z \bs\tau_j^z\bs\tau_{j+1}^z
)\, ,
\ee
where periodic boundary conditions are understood. 
We are interested in the expectation values of local (in the $\sigma$-representation) operators $\bs O$. Since the initial state is eigenstate of $\bs\Pi^z_\sigma$ and the latter commutes with the Hamiltonian, operators anticommuting with $\bs\Pi^z_\sigma$ have zero expectation value. We can therefore restrict to operators that commute with $\bs\Pi^z_\sigma\equiv\bs\Pi^x_{\tau}$. Such operators are local also in the $\tau$-representation.
Since $\ket{\Uparrow_\ell\Downarrow}_{\tau}$ and $\ket{\Downarrow_\ell\Uparrow}_{\tau}$ are macroscopically different (in the $\tau$-representation), the off-diagonal matrix elements of the aforementioned operators between $\ket{\Uparrow_\ell\Downarrow}_{\tau}$ and $\ket{\Downarrow_\ell\Uparrow}_{\tau}$ vanish. In addition, since $\ket{\Uparrow_\ell\Downarrow}_{\tau}=\bs\Pi_{\tau}^x\ket{\Downarrow_\ell\Uparrow}_{\tau}$, the diagonal matrix elements match, thus we have
\be\label{eq:EV}
\braket{\Psi(t)|\bs O|\Psi(t)}=\braket{\Downarrow_\ell\Uparrow|e^{i\tilde{\bs H} t}\bs O e^{-i\tilde{\bs H} t}|\Downarrow_\ell\Uparrow}_{\tau}\, .
\ee
We now take the thermodynamic limit by fixing the position of the flipped spin at a finite value $\ell$ and moving the boundaries to $\pm
\infty$ (for example, index $j$ in \eqref{eq:Htau} now runs from $-\infty$ to $\infty$). Eq.~\eqref{eq:EV} is the expectation value of a local operator after a quench from a domain-wall state in an integrable model, which is nowadays studied in the framework of generalised hydrodynamics (GHD)~\cite{Bertini2016Transport,Castro-Alvaredo2016Emergent}. Given the GGEs describing the system outside the lightcone, which in our case correspond to the homogeneous global quenches with initial states $\ket{\Downarrow}_\tau$ and $\ket{\Uparrow}_\tau$, GHD allows us to characterise the state inside the lightcone.
In interacting systems GHD  relies on the assumption that at large times the state becomes locally equivalent to a macro-state characterised by the (quasi)local integrals of motion. In some specific instances the GHD predictions have been confirmed by numerical simulations, but here we need to be particularly prudent because the theory is somehow blind to localised perturbations (in the domain-wall scenario they are hidden in the ambiguity of how the states are joined), which we are instead questioning if potentially relevant.  
Let us then focus on the case $J_z=0$. This corresponds to the quantum XY model, which is a noninteracting system where generalised hydrodynamics was shown to be exact~\cite{Fagotti2020, Alba2021Generalized}. The initial state is a Slater determinant (in terms of the usual Jordan-Wigner fermions $\bs c^\dag_j=\bs\tau_j^+ \prod_{n<j}\bs\tau^z_n$) and the behaviour in the limit of large time is captured by first-order generalised hydrodynamics. In the low-inhomogeneity limit the state becomes locally equivalent to a stationary state that is completely characterised by the occupation numbers $\braket{\bs n(k)}=\braket{\bs b^\dag_k \bs b_k}$ of the Bogoliubov fermions $\bs b^\dag_k$ diagonalising the Hamiltonian~\footnote{We warn the reader that the XY model is non-abelian integrable~\cite{Fagotti2014On}, so this statement is correct only if the additional charges do not play any role, like in the present situation.}. At the leading order one finds
$
\braket{\bs n(k)}_{x,t}\asymp\braket{\bs n(k)}_{x-v(k) t,0}
$, 
where $x$ is the coarse-grained position of $\bs O$ and $v(k)$ is the velocity of the quasiparticle excitation with momentum $k$. In our specific case 
$
v(k)=-4J_x J_y\sin(2k)/[(J_x+J_y)^2\cos^2 k+(J_x-J_y)^2\sin^2 k]^{1/2}
$.

\begin{figure}[!t]
\includegraphics[width=0.4\textwidth]{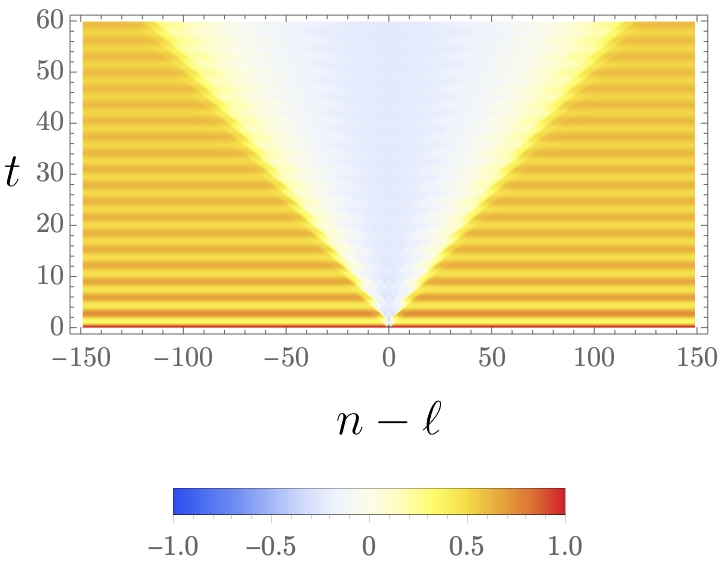}\\
\caption{Time evolution of $\braket{\bs\sigma_n^z}$ in a chain with $600$ spins with $J_x=1$, $J_y=0.5$, and $J_z=0$ from the initial state $\ket{\Uparrow\downarrow_\ell\Uparrow\downarrow_L}_\sigma$.}
\label{fig:1}
\end{figure}

\begin{figure}[!t]
\includegraphics[width=0.4\textwidth]{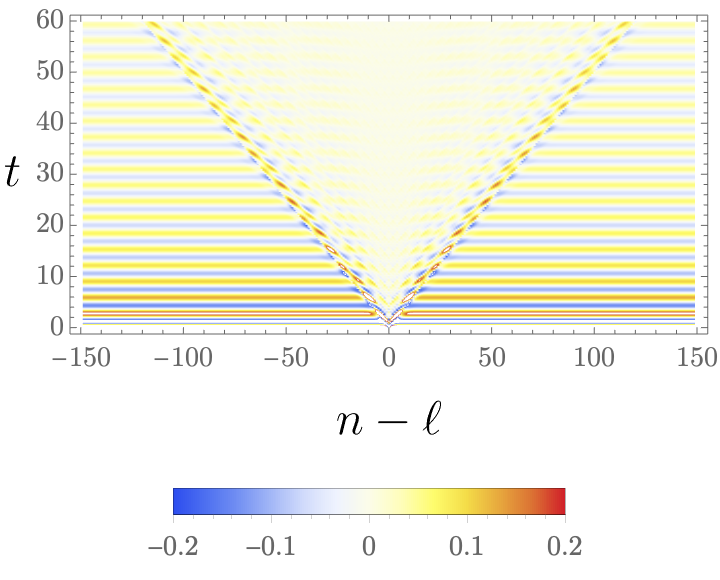}
\caption{The discrepancy between the time evolution shown in Fig.~\ref{fig:1} and the asymptotic prediction~\eqref{eq:prediction}.
The lightcone is well visible because we truncated GHD at the first order~\cite{Fagotti2020}.
}\label{fig:2}
\end{figure}

The simplest observable that is local in both representations and has a nonzero expectation value is $\bs\sigma_n^z$, which is mapped into $\bs\tau_n^z\bs\tau_{n+1}^z$.  The latter is a 4-fermion operator and its expectation value in a Gaussian state can be computed using the Wick's theorem. We finally find
\be\label{eq:predictiontau}
\braket{\Uparrow_0\Downarrow|e^{i\tilde{\bs H} t}\bs \tau_n^z\bs \tau_{n+1}^ze^{-i\tilde{\bs H} t}|\Uparrow_0\Downarrow}_{\tau}
\xrightarrow{n \sim t\rightarrow\infty}A^2(\tfrac{n}{t})-B^2(\tfrac{n}{t})
\ee
with
\be
\ba
A(\zeta)=&\int\frac{\mathrm d k}{2\pi}\frac{\theta_H(\zeta^2-v^2(k))}{1+(\frac{J_x-J_y}{J_x+J_y})^2\tan^2 k}\\
B(\zeta)=&\int\frac{\mathrm d k}{2\pi}\frac{\mathrm{sgn}(v(k))\sin(2k)\theta_H(v^2(k)-\zeta^2)}{2(\cos^2 k+(\frac{J_x-J_y}{J_x+J_y})^2\sin^2 k)^{1/2}}\, .
\ea
\ee
Consequently, in the thermodynamic limit we have
\begin{multline}\label{eq:prediction}
\braket{\Uparrow\downarrow_\ell\Uparrow|e^{i\bs H t}\bs\sigma_n^ze^{-i\bs H t}|\Uparrow\downarrow_\ell\Uparrow}_\sigma\xrightarrow{|n-\ell|\sim t\rightarrow\infty}\\
A^2(\tfrac{n-\ell}{t})-B^2(\tfrac{n-\ell}{t})\, ,
\end{multline}
which has the usual lightcone structure characterising the spreading of a localised perturbation in integrable systems~\cite{Lieb1972The,Bonnes2014Light-Cone} but, contrary to  typical cases, it does not decay in  time -- Fig.~\ref{fig:1}. Note that this is a macroscopic effect, in the sense that also the total magnetisation in an arbitrarily large subsystem is affected, provided that the time is at least comparable with the subsystem's length.
In the limit of infinite time the expectation value approaches a negative value
\be
\braket{\Uparrow\downarrow_\ell\Uparrow|e^{i\bs H t}\bs\sigma_n^ze^{-i\bs H t}|\Uparrow\downarrow_\ell\Uparrow}_{\sigma}\rightarrow-(\tfrac{J_x+J_y}{\pi \max(J_x,J_y)})^2\, .
\ee
This is different from the infinite-time limit without the initial spin flip, which instead matches the value reached outside the lightcone
\be
\braket{\Uparrow|e^{i\bs H t}\bs\sigma_n^ze^{-i\bs H t}|\Uparrow}_{\sigma}\rightarrow(\tfrac{J_x+J_y}{2\max(J_x,J_y)})^2\, :
\ee
Even if we are just interested in the behaviour at late times, the spin flip is still a relevant localised perturbation. 
Fig.~\ref{fig:2} shows the validity of the prediction.


\paragraph{Semilocal charges.}
We have provided a counterexample to the belief that localised perturbations in the initial state do not survive the limit of infinite time after a global quench; we have not yet clarified why generally this is not the case. What is so special in the model described by $\bs H$?
In this respect, the duality transformation is again helpful. We have indeed an almost perfect knowledge of $\tilde{\bs H}$ with $J_z=0$. In particular we know that the model is non-abelian integrable~\cite{Fagotti2014On,Fagotti2016Charges} and the local conservation laws are linear combinations of $\bs\tau_j^z$ and operators of the form
$\bs\tau_j^{\alpha}\prod_{m=j+1}^{j+n-1}\bs\tau_m^z \bs\tau_{j+n}^{\beta}
$, 
where $\alpha,\beta\in\{x,y\}$~\cite{Grady1982Infinite,Prosen1998A}. In order to have a local $\sigma$-representation, they should commute with $\bs\Pi^x_{\tau}$, but this happens only if they consist exclusively of operators for which $(\alpha=\beta \wedge n\text{\emph{ is odd}})\vee  (\alpha\neq \beta\wedge n\text{\emph{ is even}})$. This is realised only in a subset of the local conservation laws. In fact, such  charges are completely irrelevant in the description of the long time limit: they have the same expectation value on both sides of the junction and do not affect the emerging locally quasi-stationary state.  
We are therefore left with charges that are local only in the $ \tau$-representation. In the $\sigma$-representation they have instead the following form (in the thermodynamic limit):
\be
\bs Q=\sum\nolimits_j \bs q_j \bs\sigma_{j+1}^z\bs\sigma_{j+2}^z\cdots\, ,
\ee
where $\bs q_j$ acts like the identity everywhere except for a finite region ending at site $j$. By borrowing a terminology used in field theory, we call them `semilocal'; note, in particular, that their density $\bs q_j\prod_{n=j+1}^\infty\bs\sigma_n^z$ commutes with any local operator with support to the left of the support of $\bs q_j$ but anticommutes with any local operator to the right if the latter anticommutes with $\bs\Pi^z_{\sigma}$.
We can also define the corresponding semilocal currents, which, due to the locality of the Hamiltonian, have the same semilocal form 
$
\bs J_j[\bs q]=\bs j_j\bs\sigma_{j+1}^z\bs\sigma_{j+2}^z\cdots
$.
We believe that one could capture the long-time dynamics directly in the $\sigma$-representation by including the semilocal charges in the generalised hydrodynamic theory, but this question will be addressed in a separate investigation. 

\begin{figure}[!t]
\includegraphics[width=0.4\textwidth]{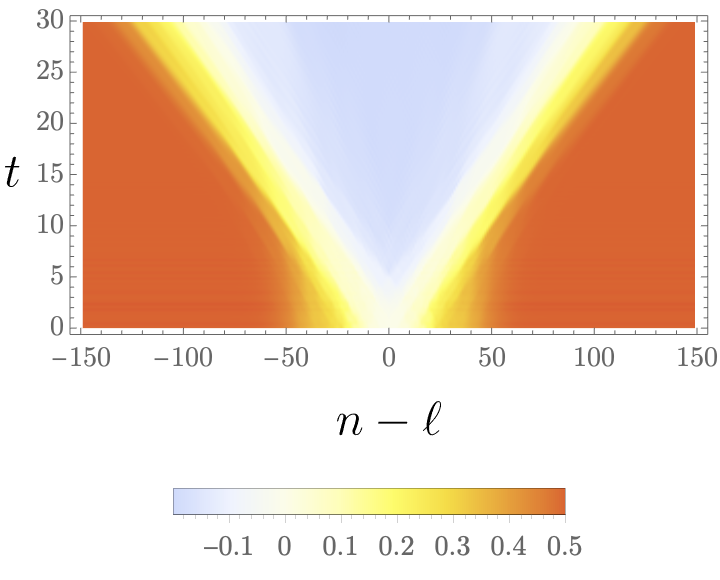}
\caption{Time evolution of $\braket{\bs\sigma_n^z}$ in a chain with $600$ spins with $J_x=J_y=1$, and $J_z=0$ after flipping the $\ell$-th spin in $e^{i\bs W}\ket{\Uparrow\downarrow_L}$, where $\bs W$ is a randomly generated translationally invariant local Hermitian operator that, in the $\tau$-representation, is noninteracting and commutes with $\bs \Pi_\tau^x$.}
\label{fig:3}
\end{figure}

\paragraph{Interaction.}
In the following we provide an argument that a similar phenomenology should be expected also in the presence of interactions. To that aim we consider the case $J_x=J_y\neq J_z$, which corresponds to the XXZ Heisenberg model. We point out that this is not a genuine global quench because the initial state is close to an eigenstate of the Hamiltonian. 
This case study is more akin to investigations on geometric quenches~\cite{Mossel2010Geometric,Alba2014Entanglement,Eisler2018Hydrodynamical,Gruber2019Magnetization,Collura2020Domain,Scopa2021Exact} and on the effects of localised perturbations in ground states with symmetry breaking~\cite{eisler2020Front,Gruber2021Entanglement,Eisler2021Entanglement} or in jammed states~\cite{Bidzhiev2021Macroscopic,Zadnik2021Measurement}. 
It is however also the arguably most studied domain-wall quench in  recent literature, so it could give an easily perceivable clue; we will discuss later how to transmute it into a global quench. 
Taking the duality transformation into account, the potential semilocal charges of $\bs H$ should be associated with charges of the XXZ model that anticommute with $\bs\Pi_{\tau}^x$. For $J_z>J_x=J_y$ only one charge of that kind is known: the component of the total spin in the direction of the anisotropy. In that case, some short-range correlators have been already studied~\cite{Piroli2017Transport}. In particular it was shown that, in the ballistic scaling limit (also known as space-time scaling limit or Euler limit) -- cf.~\eqref{eq:predictiontau} -- the sign of the magnetisation along $z$ in the states that are joined together at the initial time affects only observables odd under spin flip $z\rightarrow-z$. Since $\bs\tau_j^z\bs\tau_{j+1}^z$ is an even operator and our domain wall consists of two states that differ only in the sign of the magnetisation, we argue that the phenomenon we are describing does not present itself on the ballistic scale when $J_z>J_x=J_y$. We do not exclude it to materialise in a different scaling limit,
but we leave it as an open question, which we can also extend to generic systems with a single semilocal charge. There are instead other choices of the coupling constants that are compatible with a ballistic spreading. 
Besides the $z$ component of the total spin, indeed, an infinitely large family of charges anticommuting with $\bs\Pi_{\tau}^x$ has been identified at special values of the anisotropy, known as roots of unity ($J_z=\cos(\pi q)J_x=\cos(\pi q)J_y$, with $q$ a rational number)~\cite{Prosen2014Quasilocal,Pereira2014Exactly}. In addition, just as the XY model considered before exhibits a non-abelian structure, so does the XXZ model at those values of the coupling constants~\cite{Zadnik2016Quasilocal,Medenjak2020Isolated,Medenjak2020Rigorous}. Contrary to the case $J_z>J_x=J_y$, for $J_z<J_x=J_y$ also the observables commuting with $\bs\Pi^x_\tau$ have a nontrivial profile in the quench from the domain wall  $\ket{\Downarrow_\ell\Uparrow}_\tau$~\footnote{Note that this problem could be solved exactly using the results of Refs~\cite{Pozsgay2017Excited} and \cite{Collura2018Analytic}.}; in the end one finds a behaviour analogous to the case $J_x=J_y$ and $J_z=0$, in which the profile is qualitatively similar to that shown in Fig.~\ref{fig:1}, except for the  state not evolving outside the lightcone. 
\paragraph{Topological properties.}
The typical properties of nonequilibrium states that are locally stationary are generally fragile under global transformations of the  initial state~\footnote{For example, Refs~\cite{Fagotti2013Reduced,Collura2020How} point to the melting of order after global quenches, so localised perturbations in ground states with symmetry breaking are not expected to survive global quenches.}.  
We now show that the interacting quench considered above can instead  be easily transmuted into a global one without changing the phenomenology. To that aim, let us transform the initial ferromagnetic state $\ket{\Uparrow}_\sigma$ with a unitary operator of the form $e^{i\bs  W}$, where $\bs  W$ is a local, translationally invariant operator commuting with $\bs \Pi^z_\sigma$ and acting nontrivally on $\ket{\Uparrow}_\sigma$. 
Since the profile of local operators after the quench from $\ket{\Downarrow_\ell\Uparrow}_\tau$ is expected to be nontrivial, there must be a density $\bs q_{\mathrm{sl}}$ of a semilocal charge satisfying  $\braket{\Uparrow|\bs  q_{\mathrm{sl}}|\Uparrow}_\tau(=-\braket{\Downarrow|\bs  q_{\mathrm{sl}}|\Downarrow}_\tau)\neq 0$.  The transformation $e^{i\bs W}$ can be interpreted as time evolution under $-\bs W$ for a unit of time. A necessary condition to kill the effect would be $\braket{\Uparrow|e^{-i\bs W}\bs  q_{\mathrm{sl}}e^{i\bs W}|\Uparrow}_\tau=0$, but the expectation value of a quasilocal operator ($\bs  q_{\mathrm{sl}}$ is quasilocal in the $\tau$ representation) is a smooth function of the time, which generally vanishes at a set of times with zero measure. Even if $\braket{\Uparrow|e^{-i\bs W}\bs  q_{\mathrm{sl}}e^{i\bs W}|\Uparrow}_\tau=0$, we can  multiply $\bs W$ by a constant and obtain something different from zero. This proves that the lightcone structure is not killed by the transformation $e^{i\bs W}$, which is nonetheless sufficient to make the quench global.
This possibility is exhibited in Fig.~\ref{fig:3} for $J_x=J_y$ and $J_z=0$, which points to the stability of the qualitative picture under a change of the initial state. 
But there is much more.  

Let us consider again  the case $J_z=0$ (but arbitrary $J_x$ and $J_y$) and imagine dressing the original perturbation (spin flip) with a  unitary operator $\bs U_\ell$, acting nontrivially only around site $\ell$ and commuting with $\bs\Pi^z_{\sigma}$: $\bs\sigma_\ell^x\rightarrow \bs U_\ell \bs\sigma_\ell^x$. This results in
\be
\braket{\Psi(t)|\bs O|\Psi(t)}=\braket{\Downarrow_\ell\Uparrow|\bs U^\dag_\ell e^{i\tilde{\bs H} t}\bs O e^{-i\tilde{\bs H} t}\bs U_\ell|\Downarrow_\ell\Uparrow}_{\tau}\, .
\ee
If $\bs U_\ell$ is Gaussian in the $\tau$-representation, Ref.~\cite{Alba2021Generalized} suggests that the local perturbation is not going to affect the long time limit. In fact, it is reasonable to expect the same conclusion also for non-Gaussian $\bs U_\ell$~\cite{Gluza2019Equilibration,Murthy.Srednicki2019}. Under this assumption, we conclude that our prediction~\eqref{eq:prediction} is stable under a change of the perturbation, provided that the latter connects the sectors with a different value of $\bs \Pi^z_{\sigma}$. 
We have checked it for a randomly generated Gaussian unitary operator in the same system considered in Fig.~\ref{fig:1}. We do not show the plot just because, very short times aside, it is practically indistinguishable from Fig.~\ref{fig:1}. 
On the other hand, if we replace the spin flip resulting in \eqref{eq:Psi0} by a unitary operator $\bs U_\ell$ with the same properties as above, we end up in the $\tau$-representation with a global quench from the state with all spins in the $z$ direction after a localised perturbation. Again, this perturbation is not expected to affect the long time limit~\cite{Gluza2019Equilibration,Murthy.Srednicki2019}, so we predict the restoration of translational invariance. From this qualitative argument we argue:
\begin{multline}\label{eq:prediction1}
\braket{\Uparrow|\bs V_\ell^\dag e^{i\bs H t}\bs \sigma_n^z e^{-i\bs H t}\bs V_\ell|\Uparrow}_\sigma\xrightarrow{|n-\ell|\sim t\rightarrow\infty}\\
\tfrac{1+\delta_{\bs V}}{2}\left(\tfrac{J_x+J_y}{2\max(J_x,J_y)}\right)^2+\tfrac{1-\delta_{\bs V}}{2}[A^2(\tfrac{n-\ell}{t})-B^2(\tfrac{n-\ell}{t})]\, ,
\end{multline}
where $\bs V_\ell$ can be any unitary operator localised around site $\ell$ (in the $ \sigma$-representation) and
$
\delta_{\bs V}=\nicefrac{(\braket{\Uparrow|\bs \Pi_\sigma^z\bs V_\ell^\dag \bs \Pi_\sigma^z\bs V_\ell|\Uparrow}_\sigma+h.c.)}{2}\, .
$
Note that $\bs V_\ell$ could also represent multiple perturbations in different positions, for example $\bs V_\ell=\bs\sigma_{\ell-j}^x\bs\sigma_{\ell+j}^x$. In that case there can be multiple time scales in which lightcones emerge and merge. In the end, however, only the parity of the number of perturbations connecting the two sectors will be discriminating. 

\paragraph{Discussion.}
We have shown that, even after global quenches, localised perturbations to the initial state can have macroscopic effects.
We have proved it in a model that can be mapped to noninteracting fermions and provided evidence that  analogous behaviours can be seen in the presence of interactions preserving integrability. We traced the phenomenon back to the existence of semilocal charges, which should be incorporated in the generalised Gibbs ensemble, as well as in the generalised hydrodynamic theory. How to do that remains an open question. 
All the examples discussed in which we expect relevant localised perturbations are described by Hamiltonians with a non-abelian set of (quasi)local conservation laws. We did not use this hidden structure in the case we solved analytically, but we still wonder whether a duality transformation mapping non-abelian integrable systems into systems with (infinitely many) semilocal charges should be generally expected.

\paragraph{Acknowledgement.}
I thank Bruno Bertini, Saverio Bocini and Lenart Zadnik for discussions and Fabian Essler for insightful comments. I thank Kemal Bidzhiev for having shared some data from DMRG simulations.   

This work was supported by the European Research Council under the Starting Grant No.  805252 LoCoMacro.

\bibliographystyle{apsrev4-2}
\bibliography{references.bib}
\end{document}